\documentclass[preprint,3p,times,twocolumn,number]{elsarticle}

\usepackage{graphicx}

\usepackage{amssymb}
\usepackage{amsmath}
\usepackage{wasysym}

\usepackage{lineno}

\newcommand{\iso}[2]{$^{#1}$#2}

\usepackage{hyperref} 

\biboptions{sort&compress}

\journal{}
\begin{document}
\begin{frontmatter}
\title{First demonstration of a sub-keV electron recoil energy threshold in a liquid argon ionization chamber}
\author[llnl]{S.~Sangiorgio\corref{cor1}}
\ead{samuele@llnl.gov}
\author[llnl,berk]{T.~H.~Joshi}
\author[llnl]{A.~Bernstein}
\author[live]{J.~Coleman}
\author[penn]{M.~Foxe}
\author[llnl]{C.~Hagmann}
\author[penn]{I.~Jovanovic}
\author[llnl]{K.~Kazkaz}
\author[live]{K.~Mavrokoridis}
\author[llnl]{V.~Mozin}
\author[llnl]{S.~Pereverzev}
\author[llnl]{P.~Sorensen}

\cortext[cor1]{Corresponding author}

\address[llnl]{Lawrence Livermore National Laboratory, 7000 East Ave., Livermore, CA 94550, USA}
\address[live]{Department of Physics, University of Liverpool, Oxford St, Liverpool, L69 7Ze, UK}
\address[penn]{Department of Mechanical and Nuclear Engineering, Pennsylvania State University, University Park, PA 16802}
\address[berk]{Department of Nuclear Engineering, University of California, Berkeley, CA 94720, USA}

\begin{abstract}
We describe the first demonstration of a sub-keV electron recoil energy threshold in a dual-phase liquid argon time projection chamber. This is an important step in an effort to develop a detector capable of identifying the ionization signal resulting from nuclear recoils with energies of order a few keV and below. We obtained this result by observing the peaks in the energy spectrum at 2.82 keV and 0.27 keV, following the K- and L-shell electron capture decay of \iso{37}{Ar}, respectively. The \iso{37}{Ar} source preparation is described in detail, since it enables calibration that may also prove useful in dark matter direct detection experiments. An internally placed \iso{55}{Fe} x-ray source simultaneously provided another calibration point at 5.9 keV. We discuss the ionization yield and electron recombination in liquid argon at those three calibration energies.
\end{abstract}

\end{frontmatter}


\section{Introduction}
Dual-phase noble liquids detectors have become a popular choice for WIMP dark matter  experiments thanks to their target mass scalability, low background, and low threshold. More recently, they have been proposed also to search for coherent neutrino-nucleus scattering (CNNS).
In both applications, the only signature in the detector is a low- energy nuclear recoil. For the most common WIMP detection scenarios, nuclear recoil energies between 10-100 keVr in liquid Ar or Xe are expected \cite{Gaitskell:2004gd}.  For CNNS, detection sensitivity to nuclear recoils of a few hundred eVr is needed, especially if reactor neutrinos are employed \cite{Hagmann:2004uv}. 
Energy thresholds of 5-10 keVr are commonly obtained in dark matter experiments with liquid xenon or argon. The XENON10 experiment recently showed that an energy threshold as low as ~1 keVr, corresponding to only a handful of ionized electrons, might be obtained \cite{Angle:2011th} in liquid xenon. This result was obtained from extrapolation of the model described in \cite{Sorensen:2011bd}. It is essential to measure the low-energy response of noble gas detectors to allow their use for CNNS, and to probe broader ranges of light-mass WIMPs.

Here we report the first demonstration of sub-keV electron-recoil spectroscopy in a dual-phase argon detector. This is achieved by detecting the proportional scintillation produced in the gas phase of the detector by ionization electrons  after they have been extracted from the liquid phase. Specifically, we observed the 270 eV cascade following L-shell electron capture in $^{37}$Ar. The signal detected from spurious single electrons in the liquid is used as an absolute calibration for the ionization channel. We then suggest that low energy electron recoils can be modeled within an existing framework. 
The results shown here are a first step toward providing detector capability to measure the ionization yield of nuclear recoils at few keVr and below, as well as for enhancing the reach of axion search via the axio-electric effect.

In liquified noble gas detectors, the energy transferred to the medium by a particle results in ionization, scintillation and heat \cite{doke_absolute_2002}. The fraction lost to each channel depends on both the incident particle type and energy. Indeed, electron recombination rate varies with ionization density along the track as well as the electric field. A discussion of the mechanism and yields of energy transfer in liquefied noble gases lies outside the scope of this paper. Extensive prior work has been done on this subject (see for example \cite{Chepel:2012sj} and references therein). However, experimental data are still scarce at low energies for both electron and nuclear recoils. In addition, the measurements have mostly focused so far on scintillation yields alone, rather than ionization.

\section{Detector and calibration sources}
We have built a prototype dual-phase liquid argon time-projection chamber in the context of our program to detect coherent neutrino-nucleus scattering \cite{SangiorgioPhysProcedia2012}.
Our detector emphasizes high-sensitivity measurement of the ionization signal by means of secondary scintillation in the gas region. A particle interaction in the liquid phase produces primary scintillation (S1) and ionization \cite{doke_absolute_2002}. The electrons are drifted away from the interaction site by an electric field and extracted into the gas where they create secondary scintillation (S2) that is detected by photomultiplier tubes (PMTs). Light collection was optimized for the S2 rather than S1 signal, and as a result the detection of S1 is limited to above approximately 10 keV. 

\begin{figure}[t]
\centering
\includegraphics[width=0.75\columnwidth]{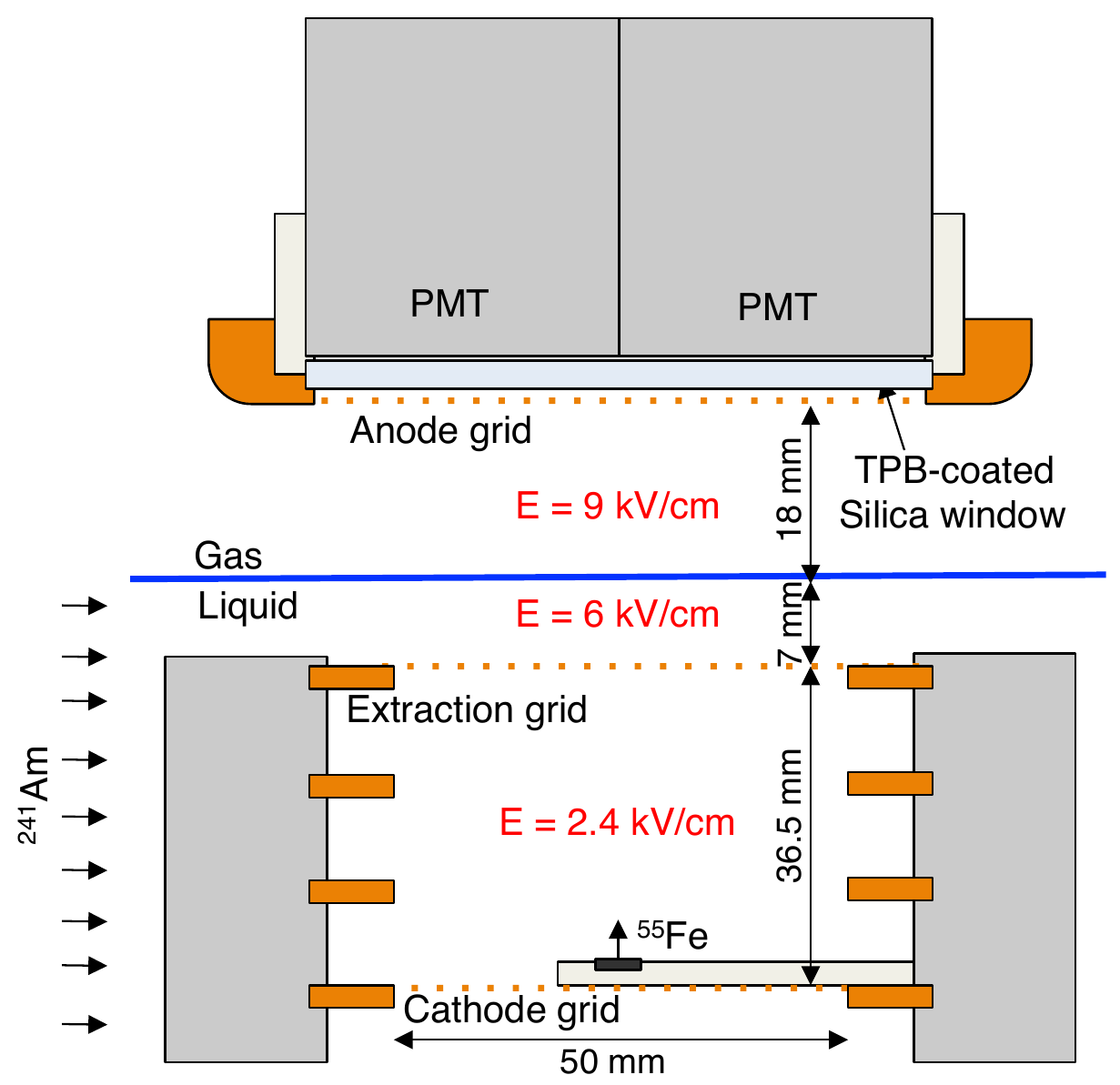}
\caption{Schematic drawing of the vertical cross section of the detector central region. The four copper rings are negatively biased and shape the electric field in the drift region. The anode is at ground potential. A SS mesh (50 mesh/inch, 30 $\mu$m dia wire, 88\% optical transparency) is used for the anode and cathode grids. The extraction grid is obtained using 15 $\mu$m SS wire with 1 mm spacing. The scintillation light is collected by four 1" PMTs through a TPB-coated silica window. The electric fields used to obtain the data reported here are also shown. For calibration, a vertical beam of gamma rays from a collimated \iso{241}{Am} source entered the active region from the side. A \iso{55}{Fe} source is mounted facing upward on a movable Teflon arm extending in the detector volume just above the cathode grid. }
\label{fig:detector} 
\end{figure}

A schematic view of the detector is provided in Fig. \ref{fig:detector}. It consists of $\sim$100 g of liquid argon as the active volume surrounded on the bottom and sides by $\sim$1 kg of inactive liquid. A gaseous region extends on top of the liquid and is kept at a constant pressure of $\sim$820 torr. The temperature of the gas is that of the Ar vapor at that pressure. The active volume is 3.65 cm high, and 5 cm in diameter. A set of four copper rings delimits this active volume and shapes the electric field needed to drift the electrons in the liquid. An extraction grid is placed 5-10 mm below the liquid level, while the anode is located 2.5 cm above the extraction grid. Electric fields up to 4 kV/cm in the drift region and up to 10 kV/cm in the gas region were achieved. At the operating fields indicated in Fig.~\ref{fig:detector}, the electrons speed is 2.85 mm/$\mu$s and 0.7 cm/$\mu$s in the drift and gas region respectively \cite{Dias:JPhysD1986, Walkowiak:NIMA2000}. The S2 signal decays with the characteristic 3.2 $\mu$s decay constant of argon triplet states in gas \cite{Keto:PhysRevLett1974}.

The argon scintillation light is transmitted through a fused silica window  and detected by four 1'' Hamamatsu R8520 PMTs modified for cryogenic operation and arranged into a square array. A 0.05 mg/cm$^2$ coating of  TetraPhenyl Butadiene (TPB) deposited on the silica window acts as wavelength shifter for the Ar UV scintillation light \cite{Boccone:2009kk}. The signals produced by PMTs provide a trigger and are digitized at 500 MHz using a LeCroy Waverunner 104MXi-A 8 bit digital oscilloscope.

The detector response was studied using the 59.54 keV gamma rays from a 5 cm diameter electroplated \iso{241}{Am} source (18.5 MBq).  To minimize peripheral events, the source was vertically collimated to produce a $\sim$2mm width beam passing through the center of the detector and spanning the entire drift region of the detector (see Fig.~\ref{fig:detector}).  The decrease in S2 amplitude of photoelectric events (59.54 keV) with event depth facilitates measurement of electron lifetime in the liquid.  The event depth is calculated using the time delay between the S1 and S2 scintillation signals. For the measurements reported here, the electron lifetime was $\sim$12 times longer than the electron drift time through the active region. The \iso{241}{Am} purity data were taken at a drift field of 3.0 kV/cm.

Two radioactive sources, \iso{55}{Fe} and \iso{37}{Ar}, were used for calibration of the detector to low energy electron recoils. A 5 mm diameter electroplated \iso{55}{Fe} bare source provided 5.90 keV and 6.49 keV x-rays with $\sim 1$ kBq activity. 
The source was mounted on a movable arm located just above the cathode inside the liquid argon active volume.

In addition, for the first time in a dual-phase detector, we used \iso{37}{Ar} as a diffused source throughout the active region. \iso{37}{Ar} was obtained from neutron irradiation of \iso{\text{nat}}{Ar} at McClellan Nuclear Research Center, similarly to what described in \cite{Aalseth:2010hx}. \iso{\text{nat}}{Ar} is composed by \iso{40}{Ar} (99.6\%), \iso{36}{Ar} (0.34\%) and \iso{38}{Ar} (0.06\%). The irradiation therefore produces mainly \iso{41}{Ar} that quickly decays with a half-life of 110 minutes, leaving \iso{37}{Ar} as the only significant radioactive product.  \iso{37}{Ar} decays by electron capture (EC) to the ground state of \iso{37}{Cl} with a half life of 35 days. The atomic shell cascade processes result in a total energy release of 2.82 keV for K-shell capture and 0.27 keV for L-shell capture \cite{Barsanov:2007fe}. For production of \iso{37}{Ar} used in our experiment, 1 liter of \iso{\text{nat}}{Ar} at a pressure of 11 bar was irradiated in a nuclear reactor for 4 hours to obtain 30 $\mu$Ci of \iso{37}{Ar}. The activity was verified by measuring gamma  emission from \iso{41}{Ar} shortly after irradiation. The gas was then cryogenically extracted, transferred to a small gas cylinder and pressurized with \iso{\text{nat}}{Ar} to 90 bar. The activated argon is injected in the detector through a purifier (SAES MC1500-903) in a similar way as the \iso{\text{nat}}{Ar}. In order to reach the desired activity in the detector, a fixed volume of gas at known pressure was introduced several time. 
We demonstrate that an \iso{37}{Ar} source is useful for calibration of low-energy electromagnetic recoils in dual-phase detectors.

\section{Results}

We report the results of our first measurements with \iso{37}{Ar}. These measurements were taken with $E_{\text{drift}}=2.4$ kV/cm applied electric field across the active region, and 6.0 kV/cm across the 7 mm of liquid argon in the extraction region, which ensures $\sim$100\% transmission of electrons from the liquid in to the gas phase \cite{Borghesani:1990}. The electric field in the gaseous amplification region was 9.0 kV/cm.

The waveform of each PMT was independently digitized and stored for off-line analysis. The trigger was set to acquire all four PMTs signals when signals from two selected PMTs exceed the hardware threshold set just above the baseline noise within 5 $\mu$s. Each waveform is 100 $\mu s$ long and includes a 35 $\mu s$ pre-trigger.  

The analysis of PMTs signals is performed following a gate-time algorithm similar to the one outlined in \cite{Kazkaz:2009zy}. From each triggered waveform, the baseline is first computed for each channel; software thresholds are then set based on the single photo-electron (s.p.e.) response of each PMT. All digitized points above threshold are grouped into a single event if their time gap is less than a given gate time across all 4 channels.  A gate time of 3.5 $\mu$s is used for the analysis of the data reported here.  Varying the gate time between 2 $\mu s$ and 7.5 $\mu s$ affects the results by $<1$\%. The single photo-electron amplitude of each PMT is used to compute the total event energy in unit of s.p.e. across all channels. 

The single photoelectron response of each PMT was calculated for each dataset by analyzing isolated single photoelectrons that appear in the tail of S2 events.  This approach determines the PMT responses at the operating temperature (89 K).

The PMTs geometric efficiency for secondary light varies by 30-40\% across the radius of the detector. In order to reject edge events, a simple fiducialization algorithm was used to quantify the X-Y position of each event. The algorithm produces a fiducial coordinate based on center-of-mass weighting of the light collected by each PMT. A fiducial cut that selects the central $\sim$40\% of the active region volume is used.

Two additional cuts are used for event selection. First of all we require that the event start time is within 5 $\mu$s from the trigger position. We also require that no more than 3 s.p.e. are present in the pre-trigger and no more than 5 s.p.e. after the end of the event. The first cut ensures that the event identified by the analysis is fully contained in the digitized waveform. The second cut limits pile-up, in particular when a low-energy event happens near the end of a high-energy event.

\begin{figure}[t]
\centering
\includegraphics[width=0.9\columnwidth]{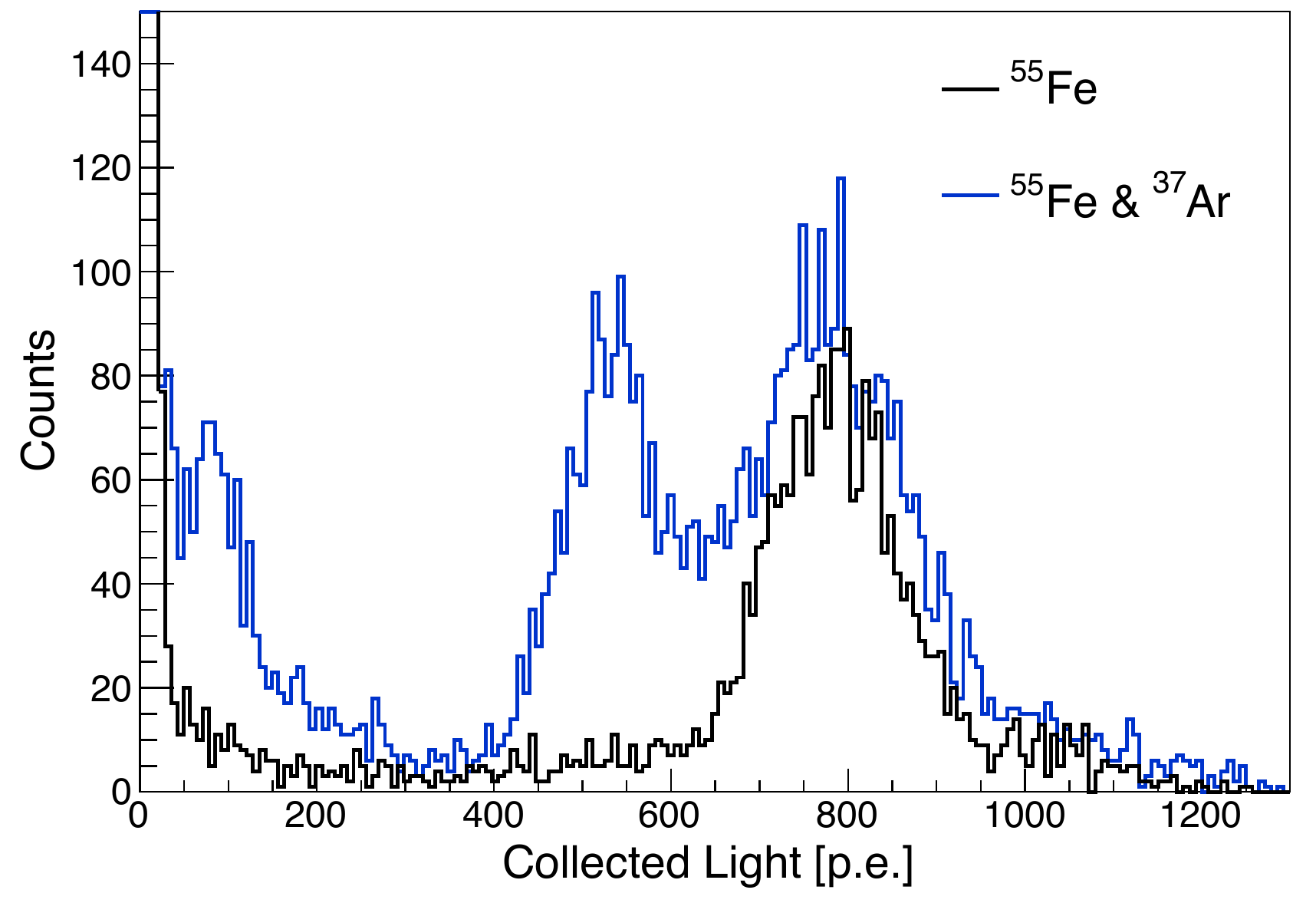}
\caption{Comparison of the spectra collected before (black line) and after (blue line) the injection of \iso{37}{Ar} in the detector. The rightmost peak present in both spectra comes from the \iso{55}{Fe} located inside the active volume. Data obtained with an electric field of 2.4 kV/cm in the drift region and of 9.0 kV/cm in the amplification region.}
\label{fig:before-vs-after-Ar37}
\end{figure}

In Fig.\ref{fig:before-vs-after-Ar37} we show the spectra collected before and after injecting $\sim$2 kBq of \iso{37}{Ar} in our detector. Before injection, only the peak from the \iso{55}{Fe} source is present. After \iso{37}{Ar} is liquefied and mixed with the rest of the argon, two more peaks appeared from K- and L-shell electron capture of \iso{37}{Ar}. 

\begin{figure}[tbp]
\centering
\includegraphics[width=0.9\columnwidth]{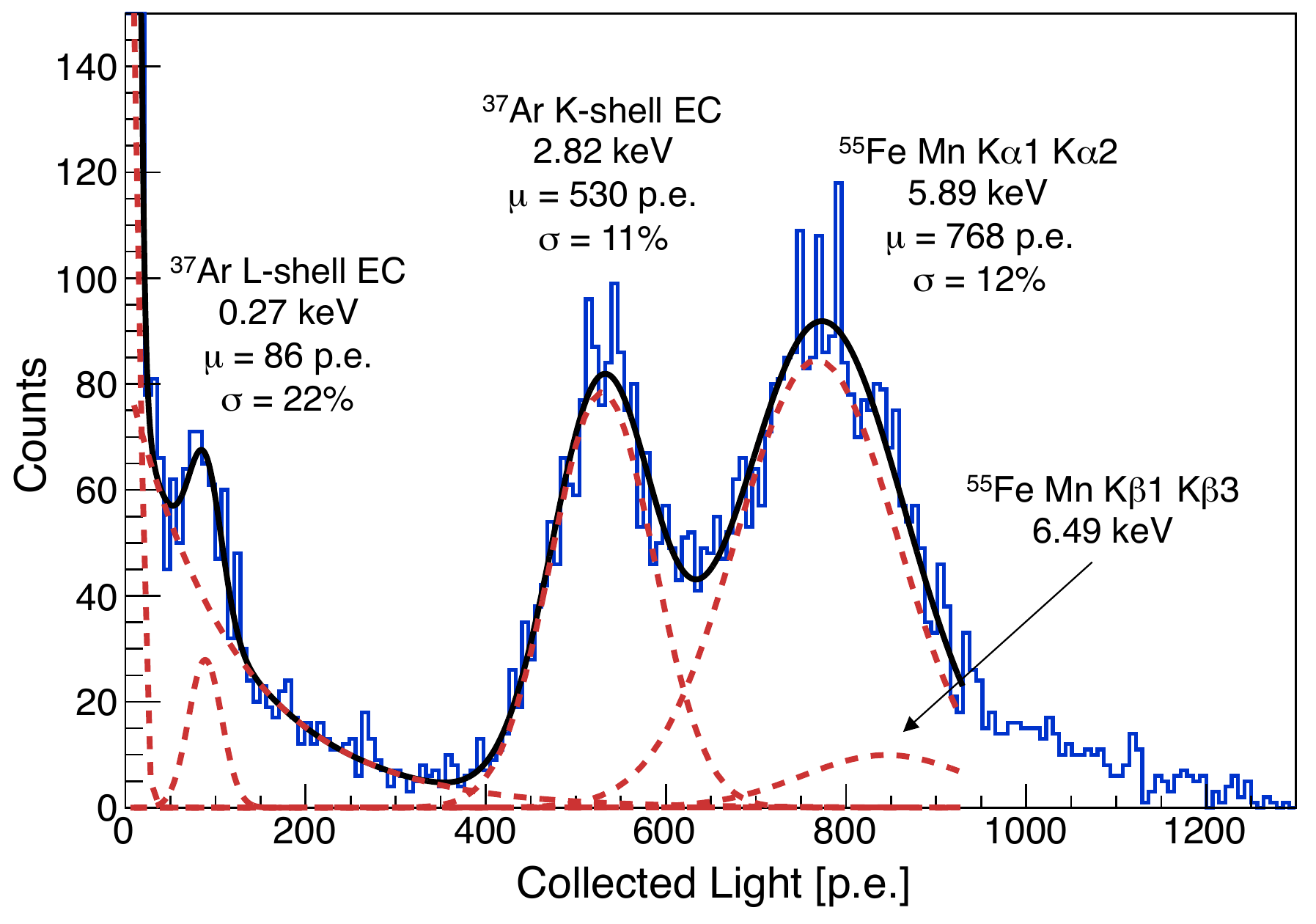}
\caption{Fit to the spectrum of the data collected after injection of \iso{37}{Ar}. The peaks are due to the \iso{55}{Fe} and \iso{37}{Ar} calibration sources. The total fit is shown in black and the individual components of the fit as red dashed lines. Data obtained with an electric field of 2.4 kV/cm in the drift region and of 9.0 kV/cm in the amplification region.  $\chi^2$/d.o.f. = 125.5/115.}
\label{fig:Ar37spectrumfit}
\end{figure}

The spectrum with \iso{37}{Ar} is fitted in Fig.\ref{fig:Ar37spectrumfit}. The low-energy background is modeled with two exponentials.  The \iso{55}{Fe} source is described by two Gaussians, one at 5.90 keV and the second at 6.49 keV. These values come from the weighted average of the four most prominent x-ray lines in the \iso{55}{Fe} decay \cite{LUND-TOI}. The relative amplitude of the two peaks was constrained accordingly to a factor $(2.99/25.4)$. The width of the smaller peak at 6.49 keV was also fixed from the value of the primary 5.90 keV peak.
The two \iso{37}{Ar} peaks were also modeled with two independent gaussians. 

A relative branching ratio of $0.116\pm0.013$ for the L- over K-shell electron capture for \iso{37}{Ar} is calculated from the fit. This branching ratio is in good agreement with previous measurements in gaseous argon
\cite{PhysRev.120.2196, Manduchi:1961, Dougan:1962, Totzek:1967}.

\begin{figure}[tbp]
\centering
\includegraphics[width=0.9\columnwidth]{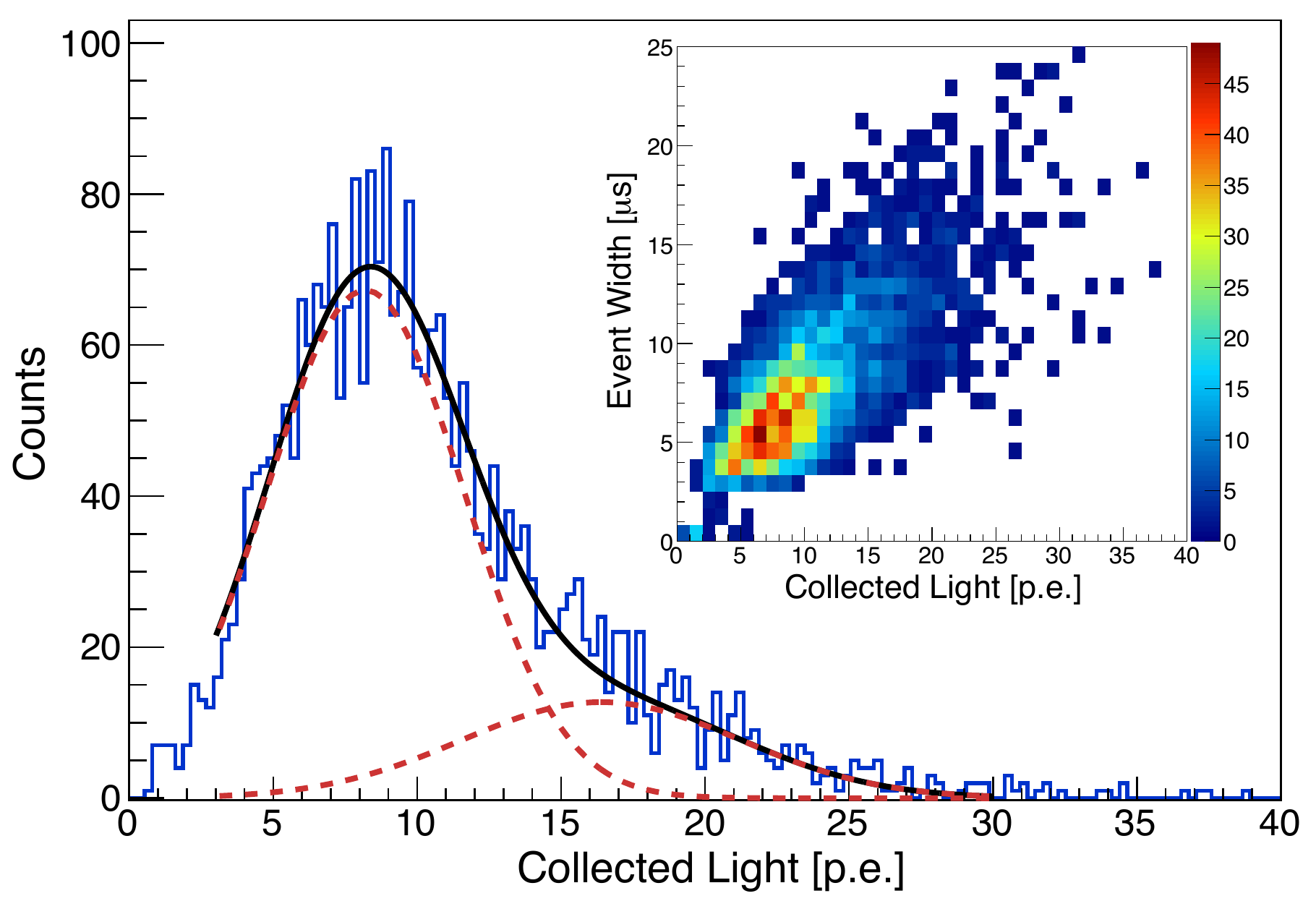}
\caption{Single electron spectrum obtained in the same electric field configuration of the \iso{37}{Ar} data. The spectrum is fitted with two gaussians describing the single- and double-electrons events. Data obtained with an electric field of 2.4 kV/cm in the drift region and of 9.0 kV/cm in the amplification region. $\chi^2$/d.o.f. = 115.4/94. In the inset, scatter plot of the event width vs the collected light in p.e. for the events in the spectrum. The distribution shown in the inset agrees with expected behavior for single and double-electron events, the extended length of some double-electron events is a result of pile-up of two single-electron events. }
\label{fig:se}
\end{figure}

During the same run we were able to measure the detector response to single ionization electrons (i.e.). Single electron events appeared following electric discharge in the detector, their rate decreasing over the course of several hours. The production mechanism of those single electrons is not fully understood at this time. The single i.e. spectrum is shown in Fig.~\ref{fig:se}. In order to study the population of single electrons, a stricter cut is applied, requiring less than 1 s.p.e. before or after each event. The fiducialization cut was removed as it is not effective at this very low number of p.e. The average event width for single electrons was $\sim$6~$\mu$s, with longer events due to pile up of more than one single-electron event, as shown in the inset of Fig.~\ref{fig:se}. The  spectrum is fitted using two gaussians to describe single- and double-electrons events. The mean and width of the double-electrons gaussian are constrained from the values of the single electrons distribution. The single electron response is 8.2 p.e/i.e.~with a 1 sigma resolution of 3.4 p.e. This number of p.e./i.e.~is consistent with the yield of secondary scintillation light in gas argon \cite{Monteiro:2008zz} and the estimated collection efficiency in our detector (approximately 1-2\%).

\begin{figure}[tbp]
\centering
\includegraphics[width=0.9\columnwidth]{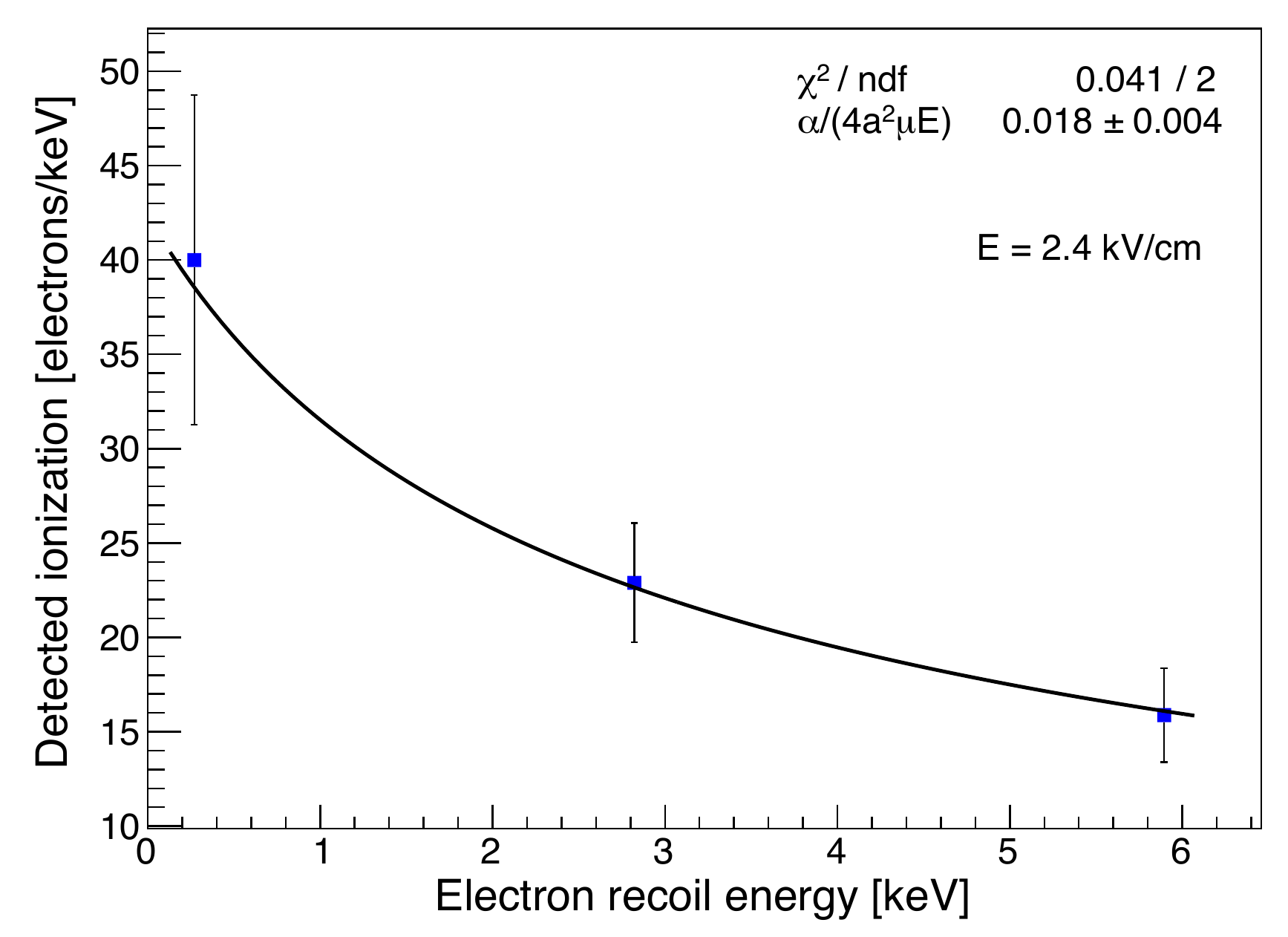}
\caption{Dependence of detected ionization in the \iso{37}{Ar} and \iso{55}{Fe} peaks from the spectrum of Fig.\ref{fig:Ar37spectrumfit} on the induced electron recoil energy. The solid curve is the result of the fit of the experimental data using the Thomas-Imel box model as described in the text.
The value of the single free fit parameter is provided in the figure. Data obtained at an electric field of 2.4 kV/cm. }
\label{fig:TI-model-fit}
\end{figure}

Using the single electron response, we compute the number of electrons extracted from liquid argon for each of the three main peaks in Fig. \ref{fig:Ar37spectrumfit}. These are plotted in Fig. \ref{fig:TI-model-fit} as a function of the nominal energy deposited in the detector. 

\section{Discussion}

It is interesting to compare the data in Fig. \ref{fig:TI-model-fit} against the Thomas--Imel box model \cite{Thomas:1987zz} that predicts the fraction of electrons that escape recombination as 

\begin{equation}
\frac{n_e}{N_i} = \frac{1}{\xi}\ln(1+\xi), \quad  \xi = \frac{N_i \alpha}{4a^2uE}
\label{eq:TI}
\end{equation}

\noindent where $\alpha$ and $u$ are the recombination and mobility coefficients, $E$ the electric field , $a$ the box dimension parameter and $n_e$ the number of electrons escaping recombination. 
Following the suggestion in   \cite{Sorensen:2011bd}, the number of initial electrons $N_i$ is calculated assuming a value $w_q=19.5$ eV for the energy required to create a quanta (either ionization or excitation) in liquid argon \cite{doke_absolute_2002} and an initial partitioning between the two channels of $N_\text{ex}/N_i=0.21$ \cite{kubota_evidence_1976}.


Equation \eqref{eq:TI} is fitted to the data with $\alpha/(4a^2uE)$ as a constant free parameter, as shown in Fig. \ref{fig:TI-model-fit}. These calculations are consistent with those obtained from argon extensions to the Noble Element Simulation Technique \cite{Szydagis:2011tk}.  Excellent agreement of the model with experimental data suggests that it is  applicable in liquid argon at low energies. An experimental verification of this behavior at different values of the drift electric field is currently being performed. 


In conclusion, we have shown that dual-phase argon ionization detectors are sensitive to sub-keV electron recoils. We have also provided a novel calibration technique in this energy range through the use of \iso{37}{Ar}. This first demonstration of sensitivity in the sub-keV to few keV range in liquid argon enables the possibility of a sensitive search for axion-like particles, via the axio-electric effect  \cite{Arisaka:2012pb}. We are also developing a technique to produce \iso{37}{Ar} with no \iso{39}{Ar} contamination. If this is successful, this calibration tool should be useful for noble-liquid detectors such as LUX \cite{Akerib:2012ys} and DarkSide \cite{Akimov:2012vv}.
Moreover, these data indicate that low-energy electron recoils in liquid argon can be modeled using a simple approach to electron recombination based on the Thomas--Imel box model. Efforts are underway to further corroborate these conclusions at different electric fields. We are also working to get experimental data on the ionization yield of low-energy nuclear recoils for the future development of both the CNNS and dark matter detectors.

\section*{Acknowledgments}
We are grateful to Dave Trombino for the use of his \iso{241}{Am} source and to Randy Hill for his engineering support. 
This work was performed under the auspices of the U.S. Department of Energy by Lawrence Livermore National Laboratory in part under Contract DE-AC52-07NA27344. Funded by Labwide LDRD. The work of T.~H.~J. was funded by the Lawrence Scholars program at LLNL and by the Department of Homeland Security under contract ARI-022. A portion of M.~F.'s research was performed under the Nuclear Forensics Graduate Fellowship Program which is sponsored by the U.S. Department of Homeland SecurityÕs Domestic Nuclear Detection Office and the U.S. Department of Defense Threat Reduction Agency.
LLNL-TR-611492.

\bibliographystyle{model1a-num-names}
\bibliography{CNS}

\end{document}